\documentclass[conference]{IEEEtran}
\IEEEoverridecommandlockouts
\usepackage{cite}
\usepackage{amsmath,amssymb,amsfonts}
\usepackage{algorithmic}
\usepackage{graphicx}
\usepackage{textcomp}
\usepackage{xcolor}
\usepackage{subcaption}
\usepackage{url}
\def\BibTeX{{\rm B\kern-.05em{\sc i\kern-.025em b}\kern-.08em
    T\kern-.1667em\lower.7ex\hbox{E}\kern-.125emX}}
\begin{document}

\title{Speech Emotion Recognition using Supervised Deep Recurrent System for Mental Health Monitoring}
%{%\footnotesize \textsuperscript{*}Note: Sub-titles are not captured in Xplore and

\author{\IEEEauthorblockN{Nelly Elsayed}
\IEEEauthorblockA{\textit{School of Information Technology} \\
\textit{University of Cincinnati}\\
OH, United States \\
elsayeny@ucmail.uc.edu}
\and
\IEEEauthorblockN{Zag ElSayed}
\IEEEauthorblockA{\textit{School of Information Technology} \\
\textit{University of Cincinnati}\\
Ohio, United States\\
elsayezs@ucmail.uc.edu}
\and
\IEEEauthorblockN{Navid Asadizanjani}
\IEEEauthorblockA{\textit{Dep. of Electrical \& Computer Engineering} \\
\textit{University of Florida}\\
Florida, United States\\
nasadi@ece.ufl.edu}
\and
\IEEEauthorblockN{Murat Ozer}
\IEEEauthorblockA{\textit{School of Information Technology} \\
\textit{University of Cincinnati}\\
Ohio, United States \\
ozermm@ucmail.uc.edu}
\and
\IEEEauthorblockN{Ahmed Abdelgawad}
\IEEEauthorblockA{\textit{School of Engineering and Technology} \\
\textit{Central Michigan University}\\
Michigan, United States \\
abdel1a@cmich.edu}
\and
\IEEEauthorblockN{Magdy Bayoumi}
\IEEEauthorblockA{\textit{Dep. of Electrical \& Computer Engineering} \\
\textit{University of Louisiana at Lafayette}\\
Louisiana, Unted States \\
magdy.bayoumi@louisiana.edu}
}
%%%%%%%%%%%%%%%%%%%%%%%%%%%%%%%%%%%%%
\thispagestyle{empty}

\begin{huge}
	IEEE Copyright Notice
\end{huge}

\vspace{5mm} %5mm vertical space

\begin{large}
	Copyright (c) 2022 IEEE
\end{large}

\vspace{5mm} %5mm vertical space

\begin{large}
	Personal use of this material is permitted. Permission from IEEE must be obtained for all other uses, in any current or future media, including reprinting/republishing this material for advertising or promotional purposes, creating new collective works, for resale or redistribution to servers or lists, or reuse of any copyrighted component of this work in other works.
\end{large}

\vspace{5mm} %5mm vertical space

\begin{large}
	\textbf{Accepted to be published in:} IEEE WFIoT-2022; 26 October - 11 November , 2022 - Yokohoma, Japan.
	https://wfiot2022.iot.ieee.org/
	
\end{large}

\vspace{5mm} %5mm vertical space
%%%%%%%%%%%%%%%%%%%%%%%%%%%%%%%%%%%%

\maketitle

\begin{abstract}
Understanding human behavior and monitoring mental health are essential to maintaining the community and society's safety. As there has been an increase in mental health problems during the COVID-19 pandemic due to uncontrolled mental health, early detection of mental issues is crucial. Nowadays, the usage of Intelligent Virtual Personal Assistants (IVA) has increased worldwide. Individuals use their voices to control these devices to fulfill requests and acquire different services. This paper proposes a novel deep learning model based on the gated recurrent neural network and convolution neural network to understand human emotion from speech to improve their IVA services and monitor their mental health. 

\end{abstract}


\section{Introduction}

Mental health is one of the crucial health aspects that must be monitored and treated for better physical health and a safer community and social life~\cite{patel2014mental}. Mental disorders cases are rising globally. According to the Institute for Health Metrics and Evaluation (IHME), the number of diagnosed individuals with one of the mental disorders globally has exceeded 1.1 billion individuals in 2016~\cite{zhou2019psychological}. According to the World Health Organization (WHO), during the first year of the COVID-19 pandemic, depression and anxiety disorders have increased by 25\% globally, especially among young people and women. Due to late or unreceived mental care, the number of related suicide has increased as well. The number of suicides has exceeded 700,000, meaning one person every 40 seconds dies by suicidal action related to a mental disorder~\cite{WhoSuicide}. Moreover, the number of mass shootings in the United States has exceeded 200 cases in less than the first half of the year~\cite{WashingtonPost}. 

Speech is the primary form of communication and emotional expression~\cite{lieberman2007evolution}. From childhood, even before being able to speak correct words, children express their emotions in their ununderstandable talks, such as their happiness and confusion. Juvenile, adults, and elderly individuals also express their emotions in their speech. All individuals express common emotions such as happy, sad, angry, happy, worry, fear, and neutral in their speech. However, different spoken languages produce differences in how these emotions are expressed in the speech tone and voice~\cite{kramsch2014language,sapir1921language}. In this paper, we focused on English as the most widely spoken language worldwide~\cite{julian2020most}. In addition, the availability of open-access data that addresses the speech emotion recognition problem is using English as the primary language.

There are several mental disorders that can be identified from individual's emotion changes~\cite{barlow1991disorders,kret2015emotion} such as depression disorder~\cite{reddy2010depression,maj2011does}, stress disorder~\cite{yehuda1997psychobiology,kessler2000posttraumatic}, and anxiety (worry/fear) disorders~\cite{craske2011anxiety,stein2008social}. Early diagnostic of mental disorders allows the individual to recieve the correct treatment and prevent sever illensses and even protect fom suisidal action~\cite{jencks1985recognition,bunney2002reducing}.

Intelligent Virtual Personal Assistants (IVA)~\cite{imrie2013virtual,gaggioli2018virtual} is s a software agent that can perform services for an individual based on processing users' questions or commands via text or voice, depending on the IVA design and purpose. The text-based interaction IVA are sometimes called chatbots, primarily when they are assessed by an online chat. The voice-based interaction IVA is also known as an intelligent voice assistant. The voice assistants can recognize the human speech and interpret its commands and questions~\cite{hoy2018alexa}. The most popular voice assistance is either embedded in smartphones such as Google's Assistant~\cite{lopez2017alexa} or a standalone device such as Amazon's Alexa~\cite{lopatovska2019talk}. 

Several studies address the effect of the IVA devices on individuals' social life~\cite{hornung2022ai}, markating~\cite{marinchak2018impact}, and social communication~\cite{chubarov2017modeling}. However, there are few studies on understanding the user behavior while using intelligent virtual personal assistant devices to improve the user experience. Yang et al.~\cite{yang2019understanding} attempted to understand how to improve the IVA user experience by investigating the relationship between perceived enjoyment, perceived usefulness, and product-related characteristics using a user survey. Coskun et al.~\cite{coskun2017understanding} also used questionnaire data to address the factors affecting IVA user experience. Venkataramanan et al.~\cite{venkataramanan2019emotion} proposed early attempts to address understanding the speech emotion to improve the user experience in voice-based devices. However, their work is implementation and computational expensive with a low accuracy rate.

%In their work, a deep one dimentional convolutional neural networks (1D-CNN) model has been proposed with . Despit of the 1D-CNN adaptability to the one dimentional data, however, it shows a significant low accuracy as it only being able to capture only the spatio features within the audio signal without the temporal information due to the 1D-CNN architecture. In addition, 

As the usage of the IVA has increased globally, in this paper, we propose a novel framework and emotion recognition model for the improvement of IVA user experience and mentoring mental health. The proposed framework employs the IVA to serve as a user assistant as well as a mental health monitoring device. The proposed framework service considers both the user request and emotion to provide a convenient service and improve the user experience. In addition, based on our investigation of the state-of-the-art speech emotion recognition models, we propose a novel deep architecture that combines the 1D-CNN and gated recurrent unit (GRU). The 1D CNN serves as a feature extractor from the audio signal. Therefore, our proposed model does not require any data preprocessing stage. The GRU preserves the temporal information within the audio data. Thus, both the spatial and the temporal information is processed efficiently in the current proposed model.

\begin{figure}%[htbp]
	\centerline{\includegraphics[width=8.5cm, height= 3.5 cm]{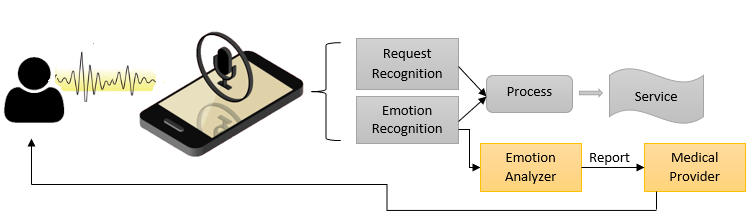}}
	\caption{The proposed emotion recogniton system for IVA service improvment and mental health monitoring.}
	\label{emotionSystem}
\end{figure}

\section{Speech Emotion Recognition}

Speech emotion recognition is one of the complex problems to solve as the emotional expression is tightly reliant on the spoken language, dialect, accent, and individuals' cultural background. In addition, the audio signal itself preserves the spatial and temporal features of the speech. There have been several attempts to solve the speech emotion recognition problem. The main two scenario approaches:
\begin{itemize}
	\item \textbf{Scenario I:} designing a model that used speech signal datasets after performing data preprocessing and feature extraction.
	\item \textbf{Scenario II:} designing a model that converts the speech signal to images (spectrograms) and then performing data preprocessing and feature extraction to fit the data to image-based models.
\end{itemize}
Wani et al.~\cite{wani2020speech} and Lotfidereshgi et al.~\cite{lotfidereshgi2017biologically} are examples of using the scenario II methodology to address the speech emotion recognition. The significant drawbacks of scenario II are that the data is processed without any consideration of temporal features in the speech signal, which significantly limits the ability of such models to recognize different emotions from the speech correctly. In addition, the data transformation and feature extraction require additional implementation costs from both hardware and software aspects. For the scenario I, Zhang et al.~\cite{zhang2013speech}, Bhargava et al.~\cite{bhargava2013improving}, Krishnan et al.~\cite{krishnan2021emotion}, and Venkataramanan et al.~\cite{venkataramanan2019emotion} proposed different machine learning and deep learning approaches to solve the speech emotion recognition under the scenario I approach. This approach requires less data transformation and feature extractions compared to the scenario II approaches. However, these models were not considering the temporal information within the audio signal within their approaches.

This paper addresses the spatial and temporal information within the speech signal. We proposed a novel model based on the 1D-CNN and the GRU to gather the spatial and temporal information within the learning approach to reduce the data preprocessing stages to fit the model within the intelligent virtual assistant devices and accelerate processing time to achieve an overall improvement of the user experience. 

\section{Proposed System}
The proposed emotion system for IVA service improvement and mental health monitoring is shown in Figure~\ref{emotionSystem}. In this system. The IVA consists of a speech recognition system and an emotion recognition system that acts together to the service production based on the user's emotion and the requested service. For example, if a user requests from the IVA to listen to a music track and at the same time the emotion recognition system determines that the user is angry, the music selection will be made to find the music track categorized as calming music to improve the user emotional experience. Another example is if the emotion recognition system determines that the user is in a fear emotion, it can advise the user to seek help. In addition, the system collects the recognized emotions through different users' requests throughout the day and provides both the user and the medical provider with the emotional statuses through the IVA using time which helps the medical provider find early alerts of mental disorders and monitor existing mental disorders treatments.

\subsection{Proposed Recognition Model}
This paper proposed a simple design hybrid model of the gated recurrent neural networks (GRU) with a 1D convolution neural network (1D-CNN) support. The GRU was first introduced by Chung et al.~\cite{chung2014empirical} as a recurrent neural network that has a simpler gated mechanism compared to the long short-term memory (LSTM), which reduces the implementation requirements from the software and hardware aspects~\cite{zaghloul2021fpga}. Similar to the LSTM and the recurrent neural network (RNN), the GRU uses its previous time step output and current input to calculate the next output~\cite{chung2014empirical}.  %The GRU architecture is shown in Figure~\ref{GRU}. 
The GRU consists of two gates: the update $z$ and reset $r$ gates. At time $t$, the output $h_t$ of the GRU can by:
\begin{align}%\begin{flalign}
		z_{t} &= \sigma (W_{xz} x_{t} + U_{hz} h_{t-1}) \label{z_gate}\\
		r_{t} &= \sigma (W_{xr} x_{t} + U_{hr} h_{t-1})\label{r_gate}\\
		{\tilde h}_{t} &= \mathrm{tanh} (W x_{t} + U( r^{(t)} \odot h_{t-1}))\label{h_hat}\\
		h_{t} &= (1-z_{t})\odot h_{t-1}+z_{t}\odot{\tilde h}_{t}\label{h}
\end{align} 
where at time step $t$, $W_{xz}$, $W_{xr}$, and $W$ are the feedforward weights of the update gate $z_{t}$, the reset gate $r_{t}$, and the output candidate activation ${\tilde h}_{t}$ .
$U_{hz}$, $U_{hr}$, $U$ are the recurrent weights are of the update gate $z_{t}$, the reset gate $r_{t}$, and the output candidate activation ${\tilde h}_{t}$, respectively. $\sigma$ is the logistic sigmoid function, $\mathrm{tanh}$ is the hyperbolic tangent function, and the symbol $\odot$ denotes the Hadamard (elementwise) multiplication. 

The GRU has successfully achieved significant results in various applications, especially in signal data such as emotion recognition from EEG signal~\cite{lew2020eeg}, sleep stage classification from EEG and EoG~\cite{sm2019sleep}, arrhythmia supraventricular premature beat detection from ECG signal~\cite{elsayed2021arrhythmia}, music source separation~\cite{liu2019dilated}, and sound event detection~\cite{lu2017bidirectional}.
%\begin{figure}%[htbp]
%	\centerline{\includegraphics[width=5cm, height=3.5 cm]{gru.png}}
%	\caption{The GRU block architecture~\cite{elsayed2018deep}.}
%	\label{gru}
%\end{figure}

\begin{figure}%[htbp]
	\centerline{\includegraphics[width=8cm, height= 14 cm]{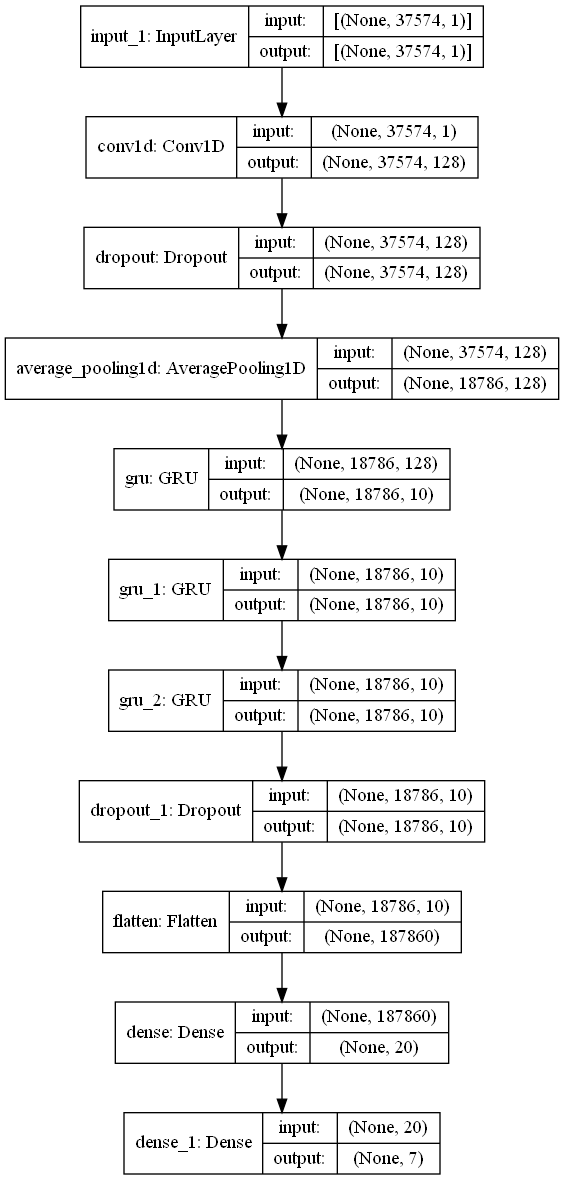}}
	\caption{The proposed GRU and 1D-CNN model for speech emotion recognition.}
	\label{full_model}
\end{figure}

GRU can learn the spatial features of the speech signal and the temporal information due to the recurrent behavior. In this paper, we selected the GRU as a competitive recurrent neural network that requires less budget and can achieve comparable results to the LSTM.
The 1D-CNN acts as the feature extractor for the 1D speech signal~\cite{kiranyaz20191,li2019feature,kiranyaz20211d}. Therefore, the significant contribution of this model is that it does not require any data preprocessing prior to the model training compared to the state-of-the-art models. The detailed architecture of the proposed model is shown in Figure~\ref{full_model}. The 1D-CNN is followed by a dropout layer with 30\% dropout rate that helps to prevent the model of overfitting problem and prevents the all the neurons in the folllowing layer from synchronously optimizing their weights~\cite{bonnin2017machine}. The average pooling layer creates a downsampling feature map and helps increase the robustness of the model~\cite{elsayed2019gated}. The flatten layer is used to adujust the input size prior to the fully connected dense layer which using a ReLU activation function~\cite{elsayed2018empirical}. Finally, a Softmax layer is used to determine the models' recognized emotion from the input speech~\cite{memisevic2010gated}.

\section{Experimental Results}

\subsection{Dataset}
\begin{figure*}%[b]
	\centering
	\begin{tabular}{cccc}
		\subcaptionbox{Angry emotion.\label{angryWave}}{\includegraphics[width = 1.5in]{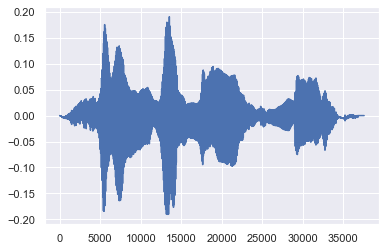}} &
		\subcaptionbox{Disgust emotion.\label{2}}{\includegraphics[width = 1.5in]{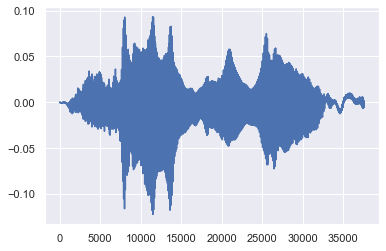}} &
		\subcaptionbox{Fear emotion.\label{3}}{\includegraphics[width = 1.5in]{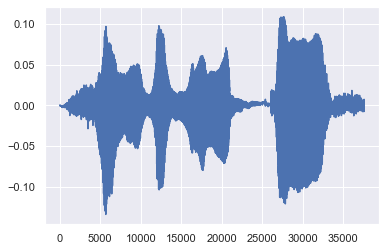}} &
		\subcaptionbox{Happiness emotion.\label{4}}{\includegraphics[width = 1.5in]{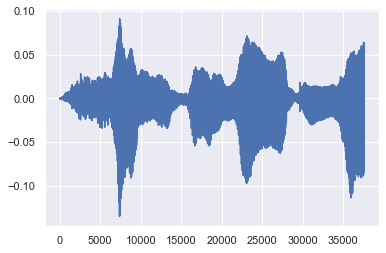}}\\
		\subcaptionbox{Pleasant/Surprise emotion.\label{5}}{\includegraphics[width = 1.5in]{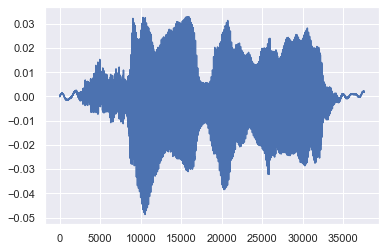}}&
		\subcaptionbox{Sadness emotion.\label{6}}{\includegraphics[width = 1.5in]{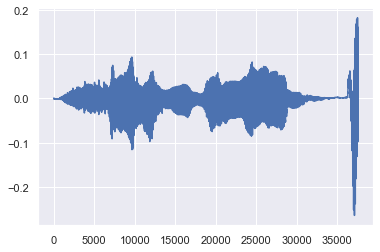}}&
		\subcaptionbox{Neutral emotion.\label{7}}{\includegraphics[width = 1.5in]{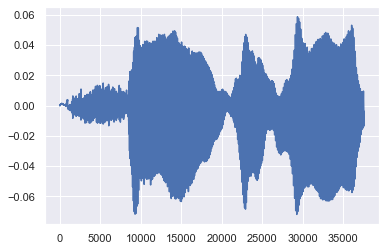}}\\
	\end{tabular}
	\caption{A sample of each category of audion emotial wave signal in the TESS dataset.}
	\label{dataSamples}
\end{figure*}

The proposed model has been empirically evaluated using the Toronto emotional speech set (TESS)~\cite{dupuis2010toronto} which is one of the emotion recognition dataset benchmarks. The speech recordings were done in the Toronto area by two actresses who speak English as their first language. The dataset consists of 2800 stimuli. The data has seven different emotion categories: anger, disgust, fear, happiness, pleasant/surprise, sadness, and neutral. The primary significance of this dataset is that the distribution between the number of stimuli per emotion category is equally likely. Figure~\ref{dataSamples} shows the waveplot for randomly selected audion from each class of the dataset.
%\begin{table}[htbp]
%	\caption{Proposed model design summary using the TESS emotion dataset}
%	\begin{center}
%		\begin{tabular}{|l|l|l|}
%			\hline
			%\textbf{Table}&\multicolumn{3}{|c|}{\textbf{Table Column Head}} \\
			%\cline{2-4} 
%			\textbf{Layer} & \textbf{\textit{ Output Shape }} &\textbf{\#Param}\\
			%\hline
			%Train Accuracy& More table copy$^{\mathrm{a}}$&  \\
%			\hline
%			InputLayer & (None, 37574, 1) & 0\\
%			Conv1D &   (None, 37574, 128)     &   512 \\
%			Dropout &  (None, 37574, 128) & 0\\
%			AvgPooling1D & (None, 18786, 128) & 0\\
%			GRU &  (None, 18786, 10)     &    4200  \\
%			GRU &    (None, 18786, 10)   &      660  \\
%			GRU&   (None, 18786, 10)  &    660 \\
%			Dropout &  (None, 18786, 10) & 0\\
%			Flatten & (None, 187860)  & 0\\
%			Dense & (None, 20) & 3757220   \\
%			Dense & (None, 7) & 147  \\
%			\hline 
			%\multicolumn{4}{l}{$^{\mathrm{a}}$Sample of a Table footnote.}
%		\end{tabular}
%		\label{tab1}
%	\end{center}
%\end{table}

\subsection{Emperical Setup Details}
This model has been implemented using Python 3.3.8, Tensorflow 2.4.0, Numpy 1.19.5, Pandas 1.2.4, and Librosa 0.9.1. Our experiments were performed on a Window 10 OS, Intel(R) Core(TM) i-9 CPU @ 3.00 GHz processor with 32-GB memory, and NVIDIA GeForce RTX 2080 Ti graphics card. 

\begin{figure}%[htbp]
	\centerline{\includegraphics[width=5 cm, height= 5 cm]{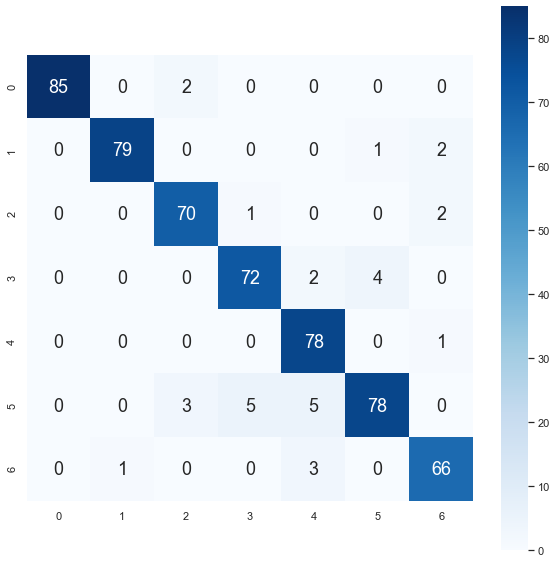}}
	\caption{The confusion matrix of the proposed model using the TESS dataset.}
	\label{confusion_matrix}
\end{figure}

The model was trained for 20 epochs, and the batch size was set to 20. The categorical cross-entropy function is used as the loss function~\cite{ketkar2017introduction}. The RMSProp has been used as the model optimization function~\cite{hinton2012neural} with learning rate $lr= 0.001$, $momentum = 0.0$, discounting factor for the history/coming gradient $rho = 0.9$, and $\epsilon = 1e-07$.
For the 1D-CNN layer, the number of kernels was set to 128 with size three. The kernel weights have been initialized using the He-uniform initializer~\cite{he2015delving} and the biases has been initialized to zero. The number of unrollments of each of the three GRU layers has been set to 10. The sigmoid $\mathit{(\sigma)}$ and hyperbolic tangent $\mathit{(tanh)}$ functions have been used as the activation and recurrent activation functions, respectively. In addition, the kernel weights of the GRU layers have been initialized using Glorot-uniform initializer~\cite{glorot2010understanding}. The biases have been initialized by zero. The 1D Average pooling layer padding has been set to valid with a pooling size of three and stride of two.

\begin{table*}%[htbp]
	\caption{A statistical analysis for the proposed model for each emotion category recogniton}
	\begin{center}
		\begin{tabular}{|l|ccccccc|}
			\hline
			\textbf{Statistical}&\multicolumn{7}{|c|}{\textbf{Emotion Category}} \\
			\cline{2-8} 
			\textbf{Analysis} & \textbf{\textit{Angry}}& \textbf{\textit{Disgust}}& \textbf{\textit{Fear}}
			& \textbf{\textit{Happiness}} & \textbf{\textit{Surprise}}& \textbf{\textit{Sadness}}& \textbf{\textit{Neutral}} \\
			\hline
			Accuracy&99.643\%    & 99.286 \%    & 98.571\%    & 97.857\%    & 98.036\%    &  96.786\%    &98.393\%\\
			F1-score& 0.98837   &   0.97531    &0.94595   &  0.92308   & 0.93413   &   0.89655   &  0.93617 \\
			Error rate&  0.00357 &  0.00714  &  0.01429 &  0.02143  &  0.01964   &  0.03214 &   0.01607  \\
			ISCI&  0.97701  &   0.95091 &   0.89224  &   0.84615 &   0.87371  &  0.79690 &  0.87243 \\  
			OP &  0.9848& 0.97527 &  0.96989 & 0.94483 &  0.97622 &  0.89626& 0.95964 \\
			Sensativity & 0.97701  &0.96341   & 0.9589   & 0.92308  & 0.98734 &  0.85714 &0.94286  \\ 
			Youden index & 0.97701 &  0.96132 & 0.94864  & 0.91063& 0.96655 &  0.84648 & 0.93265\\ 
			\hline
			%\multicolumn{4}{l}{$^{\mathrm{a}}$Sample of a Table footnote.}
		\end{tabular}
		\label{statistics_Table}
	\end{center}
\end{table*}

Table~\ref{statistics_Table} shows a statistical analysis of the proposed model result for each emotion category recognition, including the accuracy, F1-score, the area under the curve (AUC) under the ROC curve for multiple classes~\cite{hand2001simple}, the error rate, individual classification success index (ISCI)~\cite{labatut2012accuracy}, optimized precision (OP)~\cite{1688586}, sensitivity, and Youden index (Y). Figure~\ref{confusion_matrix} shows the confusion matrix of the proposed model testing where the categories anger, disgust, fear, happiness, pleasant/surprise, sadness, and neutral are indicated numerically from zero to six, respectively. The Youden index (Y) is calculated as follows:
\begin{equation}
 Y = Sensativity(\%) + Specifity (\%) - 100
\end{equation}

The overall testing results including training and testing accuracies, accuracy macro, precision, recall, F1-score, specificity~\cite{goodfellow2016deep}, the kappa value~\cite{czodrowski2014count}, and the number of trainable parameters on the proposed model are shown in Table~\ref{model_results}. The kappa value is calulated using the formula:
\begin{equation}
	\kappa = \frac{p_o - pe}{1-p_e} = 1- \frac{1-p_o}{1-p_e}
\end{equation}
\noindent
where $p_o$ is the observed positive recognition and $p_e$ is the expected positive recognition.

\begin{table}
	\caption{The proposed GRU and 1D-CNN model overall results}
	\begin{center}
		\begin{tabular}{|l|c|}
			\hline
			%\textbf{Table}&\multicolumn{3}{|c|}{\textbf{Table Column Head}} \\
			%\cline{2-4} 
			\textbf{Feature} & \textbf{\textit{Value}} \\
			\hline
			Train Accuracy & 99.107\%  \\
			Test Accuracy &  94.285\%\\
			Accuracy Macro&  98.367\%\\
			Precision&  0.94285\\
			Recall&    0.94285 \\
			F1-score& 0.94285 \\
			Specificity&  0.99047\\
			Kappa Value &0.93329\\
			No. Trainable Parameters& 3,763,399\\
			
			\hline
			%\multicolumn{4}{l}{$^{\mathrm{a}}$Sample of a Table footnote.}
		\end{tabular}
		\label{model_results}
	\end{center}
\end{table}

\begin{figure}%[htbp]
	\centerline{\includegraphics[width=6 cm, height= 3 cm]{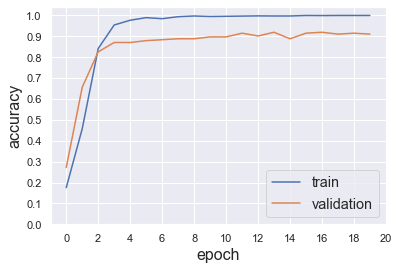}}
	\caption{The train versus validation accuracy of the proposed model over 20 epochs.}
	\label{train_accuracy}
\end{figure}

The train versus validation accuracies of the proposed model through the 20 epochs are shown in Figure~\ref{train_accuracy} where the model shows a stable training process during each epoch. 

We compared our model to the state-of-the-art speech emotion recognition models that are based on machine learning recognition methodologies. To perform a fair comparison among the different state-of-the-art and our proposed models, we selected the models that have been evaluated and trained using the TESS emotion dataset. During our other models' investigations, we found that several state-of-the-art research excluded two or more categories of the recognition task of the TESS dataset without an explicit declaration about the reasons for such reduction in the dataset that may significantly affect the model performance and purpose. Therefore, we excluded these models from our comparison due to the lack of information and uneven data usage.
Table~\ref{modelsCompare} shows the comparison between our proposed model and other models from the methodology used and the accuracy. Our model outperformed the state-of-the-art models that address the speech emotion recognition task.

\begin{table}
	\caption{Comparison between the proposed model and the state-of-the-art speech emotion recognition models}
	\begin{center}
		\begin{tabular}{|l|l|c|}
			\hline
			%\textbf{Table}&\multicolumn{3}{|c|}{\textbf{Table Column Head}} \\
			%\cline{2-4} 
			\textbf{Model} & \textbf{\textit{Method}} & \textbf{\textit{Accuracy}}\\[0.25ex]
			\hline
			Venkataramanan et al.~\cite{venkataramanan2019emotion}& 2D CNN with  & 66.000\%\\
			& Global Avg Pooling& \\
			\hline
			Sundarprasad~\cite{sundarprasad2018speech}&PCA, SVM, and& 90.000\%\\
			& Mel-Frequeny& \\
			&  Cepstrum Features& \\
			\hline
			Krishnan et al.~\cite{krishnan2021emotion}& SoA Classsifier and & 93.300\%\\
			&  Entropy features from  &\\
			& Principle IMF modes & \\
			\hline
			Lotfidereshgi et al.~\cite{lotfidereshgi2017biologically}& Liquid State Machine & 82.350\%\\
			\hline
			Zhang et al.~\cite{zhang2013speech}&Kernel Isomap & 80.850\%\\
			\hline
			Zhang et al.~\cite{zhang2013speech}&PCA & 72.350\%\\
			\hline
			Bhargava et al.~\cite{bhargava2013improving} &   Artificial Neural Nets& 80.600\%\\
			\hline
			Bhargava et al.~\cite{bhargava2013improving} &  SVM& 80.270\%\\
			\hline
			\textbf{Our Model} & GRU \& 1D-CNN  & \textbf{94.285\%}\\
			\hline
			%\multicolumn{4}{l}{$^{\mathrm{a}}$Sample of a Table footnote.}
		\end{tabular}
		\label{modelsCompare}
	\end{center}
\end{table}
\subsection{Emotion Analyzer}
The emotion analyzer stores the recognized emotion and the time of user request. Then based on the user preference settings, it builds a visual report that can be accessed through mobile and a web application, providing a flowchart of emotional changes and how frequently the emotion changes during the day. Connecting the IVA with a medical provider, the report will be sent to the medical provider for further diagnosis or monitoring for mental health disorders based on the emotions and the frequency of the emotional changes that the emotion analyzer report will provide.

\section{Conclusion}
Personal emotion is one of the most significant indicators of mental health normality and issues. In this paper, we proposed a novel deep model based on gated recurrent neural networks (GRU) and the one-dimensional convolutional neural networks (1D-CNN) to recognize different emotions from speech. The model can be applied within an intelligent virtual personal assistant to improve the user experience while combining the user request and emotion to provide the appropriate service as well as it can act as a medical health device that monitors the speech emotions and provides the medical provider with emotion changes of the patient while using the smart virtual personal assistant in their daily basics which can help to adjust medical prescriptions and further mental health issues. The significance of the proposed model is that it does not require any additional data preprocessing due to the 1DCNN that behaves to extract the features from the speech signal. Moreover, the recurrent gated unit learns the spatial and temporal features of the speech signal. Thus, the proposed model outperforms the state-of-the-art models that address the emotion recognition problem.

\bibliographystyle{ieeetr}
\bibliography{emotionReferences}
\end{document}